# EFFICIENT ALGORITHMS FOR SEVERAL CONSTRAINED ACTIVITY SCHEDULING PROBLEMS IN THE TIME AND SPACE DOMAINS

**Madalina Ecaterina Andreica**
The Bucharest Academy of Economic Studies
madalina.andreica@gmail.com

**Mugurel Ionut Andreica**
Politehnica University of Bucharest
mugurel.andreica@cs.pub.ro

**Angela Andreica**
Commercial Academy Satu Mare
academiacomerciala@yahoo.com

**Abstract:** In this paper we consider several constrained activity scheduling problems in the time and space domains, like finding activity orderings which optimize the values of several objective functions (time scheduling) or finding optimal locations where certain types of activities will take place (space scheduling). We present novel, efficient algorithmic solutions for all the considered problems, based on the dynamic programming and greedy techniques. In each case we compute exact, optimal solutions.

## 1. Introduction

Activity scheduling is an important aspect in various domains, like business processes, industrial workflows, distributed systems, and so on. Scheduling the activities efficiently can bring multiple benefits, like minimizing costs, maximizing profits and/or throughput or optimizing the social welfare of the employees. In this paper we consider several constrained time and space activity scheduling problems, for which we present efficient algorithms for computing optimal schedules. Although the considered problems are mostly tackled from a theoretical point of view, they have applications in some of the domains mentioned above, particularly those related to economic activities and computer science.

The rest of this paper is structured as follows. In Sections 2-5 we present novel algorithmic solutions for several activity scheduling problems over time. In Sections 6 and 7 we consider two space scheduling problems, where we need to find optimal locations or to divide the existing space optimally. In Section 8 we discuss related work and in Section 9 we conclude.

## 2. Scheduling K Activities over Time in order to Maximize the Total Utility

We consider a sequence of $N$ time moments. For each time moment $t$ ($1 \leq t \leq N$), a value $u(t)$ is known (which may be both positive and negative), representing the utility function if an activity is scheduled during a time interval containing $t$. We want to schedule $K$ activities during non-overlapping time intervals (i.e. consisting of disjoint time moments), such that the sum of the utilities of the time moments during which an activity is scheduled is maximum. Moreover, the time interval of the $j^{th}$ activity ($1 \leq j \leq K$) in chronological order, must contain at least $Lo(j)$ and at most $Up(j)$ time moments.

We will compute $Smax(i,j)$=the maximum sum of utilities which can be obtained by scheduling $j$ activities during the first $i$ time moments. We have $Smax(0 \leq i \leq N, 0)=0$ and $Smax(0, 1)=-\infty$ (if $Lo(1)>0$) or $0$ (if $Lo(1)=0$); $Smax(0, j>1)=-\infty$ (if $Lo(j)>0$) or $Smax(0, j-1)$ (if $Lo(j)=0$). $Smax(N, K)$ will be the maximum total utility which can be achieved.

We will first consider the case when all the upper bounds $Up(j)$ are equal to $N$ ($1 \leq j \leq K$). For each $j=1,...,K$, we will traverse, in order, all the time moments $i=1,...,N$. We will consider that we computed the partial sums $SP(*)$ ($SP(0)=0$ and $SP(1 \leq i \leq N)=SP(i-1)+u(i)$), such that we can efficiently compute $Sum(a,b)$=the sum of the utilities between the time moments $a$ and $b$ ($Sum(a,b)=SP(b)-SP(a-1)$). $Smax(i, j)=max\{Smax(i-1, j), max\{Sum(p,i)+Smax(p-1, j-1) | max\{1, i-Up(j)+1\} \leq p \leq i-Lo(j)+1\}\}$. As we traverse the time moments $i=1,...,N$ (for a fixed value of $j$), we will maintain a maximum value $Sbest$. Initially, $Sbest=Smax(0,j-1)-SP(0)$. When we reach the time moment $i$, we consider a candidate value $Scand=Smax(i-Lo(j), j-1)-SP(i-Lo(j))$ and we set $Sbest=max\{Sbest, Scand\}$. We will have $Smax(i, j) = max\{Smax(i-1, j), Sbest+SP(i)\}$. This way, the case $Up(j)=N$ (for all $1 \leq j \leq K$) can be solved in $O(N \cdot K)$ time.

In order to solve the general case, we will proceed as follows. When we compute the values $Smax(*,j)$, we will maintain a deque $DQ$, into which we will introduce $(value, moment)$ pairs. These pairs will be maintained sorted decreasingly according to the value and increasingly according to the moment. The functions $DQ.getFirst()$ and $DQ.getLast()$ ($DQ.removeFirst()$ and $DQ.removeLast()$) will be used for retrieving (removing) the first and last pair of (from) the deque (if $DQ$ is not empty). When we start computing the values $Smax(*,j)$, we will introduce into $DQ$ the first pair $(value=Smax(0,j-1)-SP(0), moment=0)$. Then, we traverse the moments $i=1,...,N$, in increasing order. When we reach a moment $i$, we perform the following actions. While $DQ.getFirst().moment<i-Up(j)$, we call $DQ.removeFirst()$. Then, we compute $Scand=Smax(i-Lo(j),j-1)-SP(i-Lo(j))$. While $DQ.getLast().value \leq Scand$, we call $DQ.removeLast()$. Afterwards, we add at the end of the deque the pair $(value=Scand, moment=i-Lo(j))$. After this, we compute $Smax(i,j)=max\{Smax(i-1,j), SP(i)+DQ.getFirst().value\}$. The time complexity is $O(N \cdot K)$ in an amortized sense.

## 3. Constrained Scheduling of K Activities over Time in order to Maximize the Total Utility

This problem is identical to the previous one, except that every activity $j$ ($1 \leq j \leq K$) must necessarily contain the special time moment $p(j)$ ($p(1)<...<p(K)$). We notice that the time moments in intervals of the form $[p(j), p(j+1)-1]$ ($1 \leq j \leq K$) can be the rightmost time moment only of activity $j$'s interval (we consider $p(K+1)=N+1$ and $p(0)=0$). We will assign $a(t)=j$ to every time moment $t$ in the interval $[left(j)=p(j), right(j)=p(j+1)-1]$ (in $O(N)$ time). Then, we can use dynamic programming and compute $Smax(i)$=the maximum sum of the utilities if the first $a(i)$ activities have been scheduled and the rightmost moment of activity $a(i)$'s time interval is smaller than or equal to $i$. We have $Smax(0 \leq i \leq p(1)-1)=0$. Then, we will compute all the values $Smax(i)$ for the time moments $i$ with the same value of $a(i)=j$ together, in increasing order of $j$ ($j=1,...,K$). Like before, we compute the partial sums $SP(*)$, with which we can evaluate $Sum(a,b)$ in $O(1)$ time.

We will first consider the case when all the values $Up(j)=N$ (i.e. there are no upper bounds). In this case, when we reach the value $j$, we will compute the values $Vmax(t)$ ($p(j-1) \leq t \leq p(j)-1$). We have $Vmax(p(j-1))=Smax(p(j-1))-SP(p(j-1))$ and $Vmax(p(j-1)+1 \leq t \leq p(j)-1)=max\{Vmax(t-1), Smax(t)-SP(t)\}$. Afterwards, we consider all the values $p(j) \leq i \leq p(j+1)-1$, in increasing order of $i$. If $i-Lo(j)+1 \leq p(j-1)$ then $Smax(i)=-\infty$ (if $i=p(j)$) or $Smax(i-1)$ (if $i>p(j)$). Otherwise, let $prev(i)=min\{i-Lo(j), p(j)-1\}$. If $i=p(j)$ then $Smax(i)=Vmax(prev(i))+SP(i)$ else $Smax(i)=max\{Vmax(prev(i))+SP(i), Smax(i-1)\}$. This case can be handled in $O(N)$ time. If, instead, we have $Up(j)=Lo(j)$ (for every $1 \leq j \leq K$), then, when considering the values $p(j) \leq i \leq p(j+1)-1$, we have: if ($i-Lo(j)+1 \leq p(j-1)$) or ($i-Lo(j)+1>p(j)$) then $Smax(i)=-\infty$ (if $i=p(j)$) or $Smax(i-1)$ (if $i>p(j)$); otherwise: if ($i=p(j)$) then $Smax(i)=Smax(i-Lo(j))+Sum(i-Lo(j)+1, i)$ else $Smax(i)=max\{Smax(i-1), Smax(i-Lo(j))+Sum(i-Lo(j)+1, i)\}$.

In order to handle the general case, we will proceed as follows. When we compute the values $Smax(i)$ ($p(j) \leq i \leq p(j+1)-1$), we will maintain a deque $DQ$, into which we will introduce *(value, moment)* pairs. These pairs will be maintained sorted decreasingly according to the value and increasingly according to the moment. We will use the functions $DQ.getFirst()$, $DQ.getLast()$, $DQ.removeFirst()$ and $DQ.removeLast()$ (defined in the previous section). Based on these functions, we define the function $DQ.insert((val, mom))$ as follows: (1) while $DQ$ is not empty and $DQ.getLast().value \leq val$ do $DQ.removeLast()$; (2) add the pair *(val, mom)* at the end of $DQ$. For each position $i$ (where $j=a(i)$), we define $tlow(i)=max\{i-Up(j), p(j-1)\}$ and $thigh(i)=min\{i-Lo(j), p(j)-1\}$. When we reach a new value of $j$, we empty the deque $DQ$. Then, we consider all the time moments $tlow(p(j)) \leq t \leq thigh(p(j))$ and call $DQ.insert((value=Smax(t)-SP(t), moment=t))$. Afterwards, if $DQ$ is empty then $Smax(p(j))=-\infty$; otherwise, we set $Smax(p(j))=DQ.getFirst().value+SP(p(j))$. For $p(j)+1 \leq i \leq p(j+1)-1$ (in increasing order of $i$), we perform the following actions: (1) for every time moment $t$ with $thigh(i-1)+1 \leq t \leq thigh(i)$ we call $DQ.insert((value=Smax(t)-SP(t), moment=t))$; (2) while $DQ$ is not empty and $DQ.getFirst().moment<tlow(i)$ do $DQ.removeFirst()$. If $DQ$ is empty, then $Smax(i)=Smax(i-1)$; otherwise, $Smax(i)=max\{Smax(i-1), DQ.getFirst().value\}$. The time complexity in this case is linear ($O(N)$) in an amortized sense.

**4. Scheduling the Largest Number of Activities**

We consider $N$ activities. Each activity $i$ ($1 \leq i \leq N$) has a fixed duration $l(i)$ and must be scheduled during $l(i)$ consecutive time moments. Moreover, each activity has a special time moment $p(i)$ which must be included within its scheduled time interval. The activities must be scheduled during non-overlapping time intervals; however, the intervals may "touch" at their endpoints, but must not intersect otherwise. Because of the constraints, it may not be possible to schedule all the activities. Thus, we want to maximize the number of scheduled activities.

We will use a greedy algorithm. First, we sort the activities in increasing order of their special time moments. Thus, we will consider that $p(1) \leq p(2) \leq .. \leq p(N)$. We will traverse the activities in this order, maintaining a stack $S$ of the activities which have been scheduled so far (the activities scheduled more recently are closer to the top of the stack). Initially, we schedule the first activity, during the interval $[p(1)-l(1), p(1)]$ (and push the activity together with its interval on the stack). When we reach the activity $i \geq 2$, we have $[x,y]$, the interval of the activity at the top of the stack. If $p(i) \geq y$, then we schedule activity $i$ in the interval $[u=max\{p(i)-l(i), y\}, v=u+l(i)]$. Then, we push the activity $i$ on the stack, together with the interval $[u,v]$ during which it was scheduled. If, instead, we have $p(i)<y$ and $l(i)<y-x$, then we remove the activity at the top of the stack $S$ (we unschedule it). Let $[x',y']$ be the interval during which the (new) activity at the top of $S$ is scheduled. This time we have $y'<p(i)$ and we can schedule activity $i$ during the interval $[u'=max\{p(i)-l(i), y'\}, v'=u'+l(i)]$; afterwards, we push the activity $i$ and the interval $[u',v']$ during which it was scheduled on top of the stack. If we have $p(i)<y$ and $l(i) \geq y-x$, then we do not schedule the activity $i$. The time complexity of this algorithm is $O(N \cdot log(N))$ for sorting the activities and $O(N)$ for traversing the activities in the sorted order and scheduling them.

**5. Lexicographically Optimal Activity Scheduling**

We consider a sequence of $N$ time moments. For each time moment $t$ ($1 \leq t \leq N$), a value $u(t)$ is known, representing the utility function if no activity is scheduled during a time interval containing $t$. We have a set of $K$ activities, each of which consists of $x$ consecutive time moments. We want to schedule the $K$ activities during non-overlapping time intervals (i.e. containing disjoint time moments), such that the chronological sequence of utilities of the time moments during which no activity is scheduled is lexicographically minimum. To be more precise, if $tm(1), ..., tm(N-K \cdot x)$ are the moments when no activity is scheduled (and $tm(i)<tm(i+1)$ for $1 \leq i \leq N-K \cdot x-1$), then the sequence $u(tm(1)), ..., u(tm(N-K \cdot x))$ is lexicographically minimum.

A simple solution is the following. We will maintain a counter $CK$ with the number of already scheduled activities and a counter $CC$ with the number of *saved* time moments (initially, $CK=CC=0$). We will also maintain a counter *pos*, meaning that all the time moments on the positions $1, ..., pos-1$ have already been considered (they are either part of a scheduled activity or are *saved*); initially, $pos=1$. While ($CK<K$) and ($CC<N-K \cdot x$), we will execute the following actions. We will select the next time moment to be saved. This is one of the moments *pos*,

$pos+x, ..., pos+i\cdot x$ ($0 \leq i \leq K-CK$). We will choose the time moment $t$ for which $u(t)$ is minimum and, in case of ties, we will choose the smallest such moment $t$. Let's assume that we selected the time moment $pos+j\cdot x$. We will increment $CK$ by $j$ (as $j$ more activities are scheduled in the intervals $[pos, pos+x-1], ..., [pos+(j-1)x, pos+j\cdot x-1]$), we will increment $CC$ by $1$ and we will set $pos=pos+j\cdot x+1$. This algorithm can be easily implemented in a time complexity of $O(N\cdot K)$. However, when $K$ is too large, this complexity is not satisfactory. We will reduce the time complexity down to $O(N)$, as follows. We will maintain a double-ended queue (deque) $DQ(r)$ for each value $r=0,1,...,x-1$. We will gradually introduce in $DQ(r)$ ($0 \leq r \leq x-1$) the utilities of the time moments $t$ (together with their associated time moments), with $t \bmod x=r$. Initially, every deque is empty. Each deque will store *(utility, moment)* pairs and provides the same functions mentioned in the previous sections. When we need to compute the minimum utility value among all the time moments $pos+i\cdot x$ ($0 \leq i \leq K-CK$), we will perform the following actions in $DQ(r=pos \bmod x)$. As long as $DQ(r)$ is not empty and $DQ(r).getFirst().moment<pos$, we will remove the first pair of the deque. If the deque is now empty, we will add in $DQ(r)$ the pair $(u(pos), pos)$ and we will set $pos'=pos+x$; otherwise, we set $pos'=DQ(r).getLast().moment+x$. While $pos' \leq pos+(K-CK)\cdot x$ do: (1) while $DQ(r)$ is not empty and $DQ(r).getLast().utility>u(pos')$, we remove the last pair from $DQ(r)$; (2) we add the pair $(u(pos'), pos')$ at the end of $DQ(r)$; (3) $pos'=pos'+x$. At the end of this loop, the first pair of $DQ(r)$ contains the smallest utility of a time moment within the required set and the associated time moment (and, in case of ties, the smallest such time moment). After finding the time moment which will be *saved*, we proceed as in the previously described algorithm. By using the deques, the amortized time complexity is $O(N)$.

## 6. Partitioning a Convex Polygon into K Vertex-Disjoint Parts with Maximum Total Area/Perimeter

In this section we consider the following partitioning problem. Given a convex polygon with $n$ vertices (numbered from $0$ to $n-1$), we want to partition it into $K$ vertex-disjoint parts with maximum total area (or weighted perimeter). Each part must be a convex polygon whose vertices are a subset of the polygon's vertices. Furthermore, no two of the $K$ parts are allowed to touch. This implies, among other things, that no two parts are allowed to share a vertex of the polygon. We also impose another condition. Each part is allowed to have at most $B \geq 0$ edges which are not also edges of the convex polygon. For $K=1$ (and any value of $B$), the optimal solution consists of the whole polygon.

For $B \geq 1$ (and any $K \geq 2$), we will compute a table $Amax(i,j,0 \leq p \leq K)$=the maximum total area (weighted perimeter) obtained if we partition the sub-polygon formed from the interval of vertices $[i,j]$ into $p$ parts. An interval of vertices $[i,j]$ is composed of the vertices $i, i+1, ..., j-1, j$ (addition and subtraction are considered modulo $n$). It should be obvious that the intervals $[i,j]$ and $[j,i]$ are different. We have $Amax(i,j,0)=0$ and $Amax(i,j,1)=A(i,j)$. We denote by $A(i,j)$ the area (weighted perimeter) of the sub-polygon composed of the interval of vertices $[i,j]$. We will pre-compute all of these values in the beginning, in $O(n^2)$ total time. We have $A(i,i)=0$, $A(i,i+1)=0$ (for area) or $w(i,i+1)$ (for perimeter) and $A(i,i+q)=A(i,i+q-1)+ATri(i,i+q-1,i+q)-(if\ (q>2)\ and\ (case=perimeter)\ then\ w(i,i+q-1)\ else\ 0)$ ($2 \leq q \leq n-1$). $ATri(a,b,c)$ denotes either the area of the triangle whose vertices are the polygon's vertices numbered $a$, $b$ and $c$, or the sum $w(a,c)+w(b,c)$ (for the perimeter case); $w(x,y)$ is the weight (e.g. length) of the segment joining the vertices $x$ and $y$ of the polygon.

In order to compute $Amax(i,j,p>1)$, we will consider several possibilities. First of all, $Amax(i+1,j,p)$ and $Amax(i,j-1,p)$ are good candidates for $Amax(i,j,p)$ (when vertex $i$ or vertex $j$ do not belong to any of the $p$ parts). The second possibility is to have vertex $i$ and vertex $j$ two vertices of two different parts. In order to do this, we consider every pair of tuples $(i, s, q)$ and $(s+1, j, p-q)$ ($i \leq s<j;\ 0 \leq q \leq p$) and compute the maximum value $TMAX(i,j,p)=max\{Amax(i,s,q)+Amax(s+1,j,p-q)\}$ over all the pairs of tuples. $TMAX(i,j,p)$ is a candidate for $Amax(i,j,p)$. The third possibility consists of having both vertices $i$ and $j$ as two vertices of the $p^{th}$ part. The $p^{th}$ part is allowed to have at most $emax=B-1$ edges which are not also edges of the polygon; when $(i,j)$ is an edge of the polygon ($j=i+n-1$), the $p^{th}$ part may have up to $emax=B$ edges which are not also edges of the polygon. We will consider every value $e$ ($1 \leq e \leq emax$) and, for each $e$, we consider every set of $e$ pairs $(a_1,b_1), (a_2,b_2), ..., (a_e,b_e)$ with the following properties: $i \leq a_1;\ b_e \leq j;\ b_l-a_l \geq 2$ ($1 \leq l \leq e$); $b_l \leq a_{l+1}$ ($1 \leq l \leq e-1$). Each pair $(a_l,b_l)$ denotes one of the edges of the $p^{th}$ part which is not also an edge of the polygon. The area (weighted perimeter) of the $p^{th}$ part, as defined by the set of $e$ pairs, is $AP(e, (a_1,b_1), ..., (a_e,b_e))=A(i,j)-(A(a_1,b_1)+...+A(a_e,b_e))$ (for the area case) or $A(i,j)-(A(a_1,b_1)+...+A(a_e,b_e))+(w(a_1,b_1)+...+w(a_e,b_e))$. For each value of $e$ and set of $e$ pairs $(a_l,b_l)$ ($1 \leq l \leq e$), we need to consider every set of $e$ numbers $q_1, q_2, ..., q_e$, with the following properties: $q_l \geq 0$ ($1 \leq l \leq e$); $q_1+q_2+...+q_l=p-1$. Then, the value $AP(e, (a_1,b_1), ..., (a_e,b_e))+Amax(a_1+1, b_1-1, q_1)+...+Amax(a_e+1,b_e-1,q_e)$ is a candidate value for $Amax(i,j,p)$. We will set $Amax(i,j,p)$ to the maximum of all the candidate values (or $-\infty$ if no candidate value exists). The optimal value of the total area of the $K$ vertex-disjoint parts is $max\{Amax(i,j,K)\}$ and the time complexity of this approach is $O(n^{max\{2\cdot B+2,3\}} \cdot K^{max\{B+2,2\}})$.

## 7. Maximum Utility Rectangular Submatrix with a Bounded Number of Distinct Heights

We have a terrain modelled as an $M$-by-$N$ matrix $A$ (with $M \leq N$). Each value of the matrix represents the height of the corresponding terrain zone. Moreover, for each position $(i,j)$ in the matrix we have a utility value $u(i,j) \geq 0$. We want to find a rectangular submatrix $B$ containing at most $K \leq M \cdot N$ distinct values (i.e. $K$ different heights), such that the aggregate (*sum* or *max*) of the utility values of the positions in the submatrix $B$ is maximum.

We will consider every row $LS=1,...,M$ as a possible upper row for the submatrix $B$. For each value of $LS$ we will create a list $List(c)$ for every column $c$ ($1 \leq c \leq N$). Initially, these lists will be empty. Then, we will consider, one at a time, every row $LJ=LS, LS+1, ..., M$ as a possible lower row for the submatrix $B$. Once the row $LJ$ is also fixed, we will traverse all the columns $c=1,...,N$ and we will add the element *A(LJ, c)* to the list *List(c)*. For every element added to a list *List(c)*, we will also maintain a counter with the number of occurrences of this element in *List(c)* (e.g. by using a hash table *HT(c)* associated to each column, where the keys are the elements' values and the values are the number of occurrences of the corresponding key). If, when adding a new element to *List(c)*, this element has never occurred before in *List(c)*, then its counter will be set to *1*; otherwise, its counter will be incremented by *1*. Thus, *List(c)* will contain all the distinct elements on the column c, between the rows *LS* and *LJ*. If *|List(c)|>K* (*|List(c)|* denotes the number of elements in *List(c)*), we will add no more element to *List(c)*. Thus, the maximum number of elements in a list *List(c)* is bounded by *min{M, K+1}*. We will now traverse the columns from left to right, maintaining two pointers, *CS* and *CD*. We initialize *CS=1* and *CD=0*. We will also maintain a list *L* with the distinct elements (and a hash table *H* with their numbers of occurrences) between the rows *LS* and *LJ* and the columns *CS* and *CD*. Initially, *L* (and *H*) will be empty. At every step *i* (*i=1,...,N*) we increment *CD* by *1* and add the elements in *List(CD)* to the list *L*; if an element *x* in *List(CD)* was not part of *L*, then we add it to *L* and set its number of occurrences (in *H*) to *1*; otherwise, we increment the number of occurrences of the element *x* (in *H*). Then, while *|L|>K*, we will perform the following steps: (1) we delete from *L* the elements *x* in *List(CS)*; if the number of occurrences of *x* (in *H*) is greater than *1*, we decrement this number by *1*; otherwise, we remove *x* from *L* (and from *H*); (2) *CS=CS+1*. At the end of each step, if *CS≤CD*, then we have a submatrix *B* with at most *K* distinct elements, with the upper row *LS*, lower row *LJ*, left column *CS* and right column *CD*. We will compute the aggregate *Bagg* of the utilities of the submatrix *B* in *O(1)* time. For the *sum* aggregate function, we can use 4 prefix sum queries (see [5]) and for the *max* aggregate function, we can use multidimensional RMQ [5]. If all the utility values are *1*, then *Bagg* is the area of the submatrix: *Bagg=(LJ-LS+1)·(CD-CS+1)*. We will set *MaxAgg=max{MaxAgg, Bagg}* (where *MaxAgg*=the maximum aggregate value found so far; initially, *MaxAgg=0*). Let's analyze the time complexity of the presented algorithm. There are $O(M^2N)$ insertion operations into the lists *List(*)*. An insertion can be performed in *O(1)* time (if we use a normal linked list and a hash table for the number of occurrences and for maintaining pointers to the location of each element *x* in the list), or in *O(log(min{M,K}))* time if we use a balanced tree (both for the list and the number of occurrences). Then, we have $O(M^2 N \cdot min\{M,K\})$ addition and/or removal operations to/from the list *L*. Again, if we use a standard linked list (for *L*) together with a hash table (*H*) for the number of occurrences and for maintaining pointers to the locations of the elements *x* in *L*, the time complexity per operation is *O(1)*. If we implement *L* as a balanced tree (which we use both as a "list" and for maintaining the number of occurrences), the time complexity is *O(log(K))* (because the list *L* never contains more than *2·K+1* distinct elements). Thus, the best time complexity that we can achieve with the presented algorithm is $O(M^2 N \cdot min\{M,K\})$.

## 8. Related Work

Activity scheduling problems have been considered in many papers. Problems regarding personnel activity scheduling in multiple domains were considered in [1, 3, 4, 6, 7, 10]. Pedestrian-route and activity scheduling theory and models were presented in [2]. Several greedy and dynamic programming algorithms for data transfer scheduling were presented in [8] and some efficient data structures were developed in [5, 9].

## 9. Conclusions and Future Work

In this paper we considered several constrained activity scheduling problems in the time and space domains. For each problem we presented novel, efficient algorithmic solutions which compute optimal schedules. As future work, we intend to consider activity scheduling problems with more complex constraints, which will have more direct applications in practical settings.